\documentstyle[sprocl]{article}
\def\be{\begin{equation}}
\def\ee{\end{equation}}
\def\bea{\begin{eqnarray}}
\def\eea{\end{eqnarray}}

\begin{document}
\title{HOW NEUTRINOS GET MASS AND WHAT OTHER\\ 
THINGS MAY HAPPEN BESIDES OSCILLATIONS}
\author{ERNEST MA}
\address{Department of Physics, University of California\\
Riverside, CA 92521, USA}
\maketitle\abstracts{In this talk I address the theoretical issue of what 
new physics is required to make $m_\nu \neq 0$.  I then discuss what other 
things may happen besides neutrino oscillations.  In particular I consider 
a possible new scenario of leptogenesis in $R$ parity nonconserving 
supersymmetry.}

\section{Introduction}

In the minimal standard model, under the gauge group $SU(3)_C \times SU(2)_L 
\times U(1)_Y$, the leptons transform as:
\begin{equation}
\left( \begin{array} {c} \nu_i \\ l_i \end{array} \right)_L \sim (1,2,-1/2), 
~~~ l_{iR} \sim (1,1,-1),
\end{equation}
and the one Higgs doublet transforms as:
\begin{equation}
\Phi = \left( \begin{array} {c} \phi^+ \\ \phi^0 \end{array} \right) \sim 
(1,2,1/2).
\end{equation}
Without additional particles at or below the electroweak energy scale, 
{\it i.e.} $10^2$ GeV, $m_\nu$ must come from the following effective 
dimension-5 operator,\cite{1}
\begin{equation}
{1 \over \Lambda} (\nu_i \phi^0 - l_i \phi^+)(\nu_j \phi^0 - l_j \phi^+).
\end{equation}
All theoretical models of neutrino mass differ only in its specific 
realization.\cite{2}

\section{Canonical, Minimal, and Next-to-Minimal Seesaw}

Add 3 heavy singlet right-handed neutrinos to the minimal standard model: 
1 $\nu_R$ for each $\nu_L$.  Then the operator of Eq.~(3) is realized 
because each heavy $\nu_R$ is linked to $\nu_L \phi^0$ with a Yukawa 
coupling $f$; and since $\nu_R$ is allowed to have a large Majorana mass 
$M_R$, the famous seesaw realtionship $m_\nu = m_D^2/M_R$ is 
obtained,\cite{3} where $m_D = f \langle \phi^0 \rangle$.  This 
mechanism dominates the literature and is usually implied when a 
particular pattern of neutrino mass and mixing is proposed.

Actually, it is not necessary to have 3 $\nu_R$'s to get 3 nonzero 
neutrino masses.  Add just 1 $\nu_R$.  Then only 1 linear combination 
of $\nu_e, \nu_\mu, \nu_\tau$ gets a seesaw mass.  The other 2 neutrino 
masses are zero at tree level, but since there is in general no more 
symmetry to protect their masslessness, they must become massive through 
radiative corrections.  As it turns out, this happens in two loops through 
double $W$ exchange and the result\cite{4} is doubly suppressed by the 
charged-lepton masses.  Hence it is not a realistic representation of 
the present data for neutrino oscillations.

Add 1 $\nu_R$ and 1 extra Higgs doublet.\cite{5}  Then 1 neutrino gets a 
seesaw mass.  Another gets a one-loop mass through its coupling to $\phi_2^0$, 
where $\langle \phi_2^0 \rangle = 0$.  This second mass is proportional 
to the coupling of the term $(\bar \phi_2^0 \phi_1^0)^2$ times $\langle 
\phi_1^0 \rangle^2$ divided by $M_R$.  The third neutrino gets a two-loop 
mass as in the minimal case.  This scheme is able to fit the present data.

\section{Heavy Higgs Triplet}

Add 1 heavy Higgs triplet $(\xi^{++}, \xi^+, \xi^0)$.  Then the dimension-4 
term
\begin{equation}
\nu_i \nu_j \xi^0 - \left( {\nu_i l_j + l_i \nu_j \over \sqrt 2} \right) 
\xi^+ + l_i l_j \xi^{++}
\end{equation}
is present, and $m_\nu \propto \langle \xi^0 \rangle$.  If $m_\xi \sim 
10^2$ GeV, this would require extreme fine tuning to make $\langle \xi^0 
\rangle$ small.\cite{6}  But if $m_\xi >> 10^2$ GeV, the dimension-4 term 
should be integrated out, and again only the dimension-5 term
\begin{equation}
(\nu_i \phi^0 - l_i \phi^+)(\nu_j \phi^0 - l_j \phi^+) = 
 \nu_i \nu_j (\phi^0 \phi^0) - (\nu_i l_j + l_i \nu_j)(\phi^0 \phi^+) + 
l_i l_j (\phi^+ \phi^+),
\end{equation}
remains, so that\cite{7}
\begin{equation}
m_\nu = {2 f \mu \langle \phi^0 \rangle^2 \over m_\xi^2},
\end{equation}
where $f$ and $\mu$ are the couplings of the terms $\nu_i \nu_j 
\xi^0$ and $\phi^0 \phi^0 \bar \xi^0$ respectively. 
This shows the interesting result that $\xi$ has a very small vacuum 
expectation value inversely proportional to the square of its mass,\cite{8}
\begin{equation}
\langle \xi^0 \rangle = {\mu \langle \phi^0 \rangle^2 \over m_\xi^2} << m_\xi.
\end{equation}
The $SU(2)_L \times SU(2)_R \times U(1)_{B-L}$ version of this relationship 
is $v_L \sim \langle \phi^0 \rangle^2/v_R$.\cite{9}

\section{Some Generic Consequences}

Once neutrinos have mass and mix with one another, the 
radiative decay $\nu_2 \to \nu_1 \gamma$ happens in all models, but is 
usually harmless as long as $m_\nu <$ few eV, in which case it will have 
an extremely long lifetime, many many orders of magnitude greater than the 
age of the Universe.  The present astrophysical limit\cite{10} is $10^{14}$ 
years.

The analogous radiative decay $\mu \to e \gamma$ also happens 
in all models, but is only a constraint for some models where $m_\nu$ is 
radiative in origin.  The present experimental limit\cite{11} on this 
branching fraction is $1.2 \times 10^{-11}$.

Neutrinoless double $\beta$ decay occurs, but is 
sensitive only to the $\nu_e - \nu_e$ entry of ${\cal M}_\nu$, which may 
be assumed to be zero in many models.  The present experimental 
limit\cite{12} is 0.2 eV.

\section{Leptogenesis in the 2 Simplest Models of Neutrino Mass}

Leptogenesis is possible in either the canonical seesaw or Higgs triplet 
models of neutrino mass.  In the canonical seesaw scenario, $\nu_R$ 
may decay into both $l^- \phi^+$ and $l^+ \phi^-$.  In the Higgs triplet 
scenario, $\xi^{++}$ may decay into both $l^+ l^+$ and $\phi^+ 
\phi^+$.  The lepton asymmetry thus generated may be converted into 
the present observed baryon asymmetry of the Universe through the 
electroweak sphalerons.\cite{13}

The decay amplitude of $\nu_R$ into $l^- \phi^+$ is the sum of tree-level 
and one-loop contributions, where the intermediate state $l^+ \phi^-$ may 
appear as a vertex correction through $\nu'_R$ exchange.\cite{14} 
The interference between them allows a decay asymmetry of $l^- \phi^+ - 
l^+ \phi^-$ to be produced, provided that $CP$ is violated.  This requires 
$\nu'_R \neq \nu_R$ and is analogous to having direct $CP$ violation in $K$ 
decay, {\it i.e.} $\epsilon' \neq 0$.

There is also $CP$ violation in the self-energy correction\cite{15} to the 
mass matrix spanning $\nu_R$ and $\nu'_R$, which is analogous to having 
indirect $CP$ violation in the $K^0 - \bar K^0$ system, {\it i.e.} $\epsilon 
\neq 0$.  This effect has a $(m - m')^{-1}$ enhancement, but the limit 
$m'=m$ is not singular.\cite{16}

Similarly, the decay amplitude of $\xi^{++}$ into $l^+ l^+$ has a self-energy 
(but no vertex) correction involving the intermediate state $\phi^+ \phi^+$. 
This generates a decay asymmetry given by\cite{8}
\begin{equation}
\delta_i \simeq {Im[\mu_1 \mu_2^* \sum_{k,l} f_{1kl} f_{2kl}^*] \over 8 \pi^2 
(M_1^2-M_2^2)} \left( {M_i \over \Gamma_i} \right).
\end{equation}
Again, $CP$ violation requires 2 different $\xi$'s.

\section{Radiative Neutrino Mass}

The generic expression of a Majorana neutrino mass is given by
\begin{equation}
m_\nu \sim f^2 \langle \phi^0 \rangle^2 / \Lambda,
\end{equation}
hence
\begin{equation}
\Lambda > 10^{13} {\rm GeV} (1~{\rm eV}/m_\nu) f^2,
\end{equation}
{\it i.e.} the scale of lepton number violation is very large (and directly 
unobservable) unless $f < 10^{-5}$ or so.

If $m_\nu$ is radiative in origin, $f$ is suppressed first by the loop 
factor of $(4 \pi)^{-1}$, then by other naturally occurring factors such 
as $m_l/M_W$ or $m_q/M_W$.  In that case, $\Lambda$ may be small enough 
to be observable directly (or indirectly through lepton flavor violating 
processes.)

Take for example the Zee model,\cite{17} which adds to the minimal standard 
model 1 extra Higgs doublet $\Phi_2$ and 1 charged singlet $\chi^+$.  Then 
the coexistence of the terms $g_{ij}(\nu_i l_j - \nu_j l_i) \chi^+$ and 
$\mu (\phi_1^+ \phi_2^0 - \phi_2^+ \phi_1^0) \chi^-$ allows the following 
radiative mass matrix to be obtained:
\begin{equation}
{\cal M}_\nu = \left[ \begin{array}{c@{\quad}c@{\quad}c} 0 & f_{\mu e} 
(m_\mu^2-m_e^2) & f_{\tau e} (m_\tau^2-m_e^2) \\ f_{\mu e} (m_\mu^2-m_e^2) 
& 0 & f_{\tau \mu} (m_\tau^2-m_\mu^2) \\ f_{\tau e} (m_\tau^2-m_e^2) & 
f_{\tau \mu} (m_\tau^2-m_\mu^2) & 0 \end{array} \right],
\end{equation}
where
\begin{equation}
f_{ij} \sim {g_{ij} \over 16 \pi^2} {\mu \langle \phi_2^0 \rangle \over 
\langle \phi^0_1 \rangle m_\chi^2}.
\end{equation}
This model has been revived in recent years and may be used to fit the 
neutrino-oscillation data.

In the above, the mass of the charged scalar $\chi$ may be light enough to 
allow observable contributions to $\Gamma (\mu \to e \nu \bar \nu)$ at tree 
level, and to $\Gamma (\mu \to e e e)$ in one loop.  Hence lepton flavor 
violating processes may reveal the presence of such a new particle. 

\section{R Parity Nonconserving Supersymmetry}

In the minimal supersymmetric standard model, $R \equiv (-1)^{3B+L+2J}$ is 
assumed conserved so that the superpotential is given by
\begin{equation}
W = \mu H_1 H_2 + f_{ij}^e H_1 L_i e^c_j + f^d_{ij} H_1 Q_i d^c_j + 
f^u_{ij} H_2 Q_i u^c_j,
\end{equation}
where $L_i$ and $Q_i$ are the usual lepton and quark doublets, and
\begin{equation}
H_1 = (h_1^0, h_1^-), ~~~ H_2 = (h_2^+, h_2^0)
\end{equation}
are the 2 Higgs doublets. If only $B$ is assumed to be conserved but not 
$L$, then the superpotential also contains the terms
\begin{equation}
\mu_i L_i H_2 + \lambda_{ijk} L_i L_j e_k^c + \lambda'_{ijk} L_i Q_j d_k^c,
\end{equation}
and violates $R$.  As a result, a radiative 
neutrino mass $m_\nu \simeq \lambda'^2 (A m_b^2)/16 \pi^2 m_{\tilde b}^2$ 
may be obtained.\cite{18}  Furthermore, from the mixing of $\nu_i$ 
with the neutralino mass matrix through the bilinear term $L_i H_2$ and 
the induced vacuum expectation value of $\tilde \nu_i$, a tree-level 
mass $m_\nu \simeq (\mu_i/\mu - \langle \tilde \nu_i \rangle /\langle 
h_1^0 \rangle )^2 m_{eff}$ is also obtained.\cite{19}

\section{Leptogenesis from R Parity Nonconservation}

Whereas lepton-number violating trilinear 
couplings are able to generate neutrino masses radiatively, they 
also wash out any preexisting $B$ or $L$ asymmetry during the electroweak 
phase transition.\cite{20,21}  On the other hand, successful leptogenesis 
may still be possible as shown recently.\cite{22}

Assume the lightest and 2nd lightest supersymmetric particles to be
\begin{equation}
\tilde W'_3 = \tilde W_3 - \epsilon \tilde B, ~~~ \tilde B' = \tilde B + 
\epsilon \tilde W_3,
\end{equation}
respectively, where $\tilde W_3$ and $\tilde B$ are the $SU(2)$ and $U(1)$ 
neutral gauginos, and $\epsilon$ is a very small number.  Note that 
$\tilde B$ couples to $\bar \tau_L^c \tilde \tau_L^c$ but $\tilde W_3$ 
does not, because $\tau_L^c$ is trivial under $SU(2)$.  Assume $\tilde \tau_L 
- h^-$ mixing to be negligible but $\tilde \tau_L^c - h^+$ mixing to be 
significant and denoted by $\xi$.  Obviously, $\tilde \tau$ may be repalced 
by $\tilde \mu$ or $\tilde e$ in this discussion.

Given the above assumptions, $\tilde B'$ decays into $\tau^\mp h^\pm$ 
through $\xi$, whereas $\tilde W'_3$ decays (also into $\tau^\mp h^\pm$) 
are further suppressed by $\epsilon$.  This allows $\tilde W'_3$ decay to 
be slow enough to be out of equilibrium with the expansion of the Universe 
at a temperature $\sim$ 2 TeV, and yet have a large enough asymmetry 
$(\tau^- h^+ - \tau^+ h^-)$ in its decay to obtain $n_B/n_\gamma \sim 
10^{-10}$.  See Figure 1.

This unique scenario requires $\tilde W'_3$ to be lighter than $\tilde B'$ 
and that both be a few TeV in mass so that the electroweak sphalerons are 
still very effective in converting the $L$ asymmetry into a $B$ asymmetry.  
It also requires very small mixing between $\tilde \tau_L$ with $h^-$, 
which is consistent with the smallness of the neutrino mass required in 
the phenomenology of neutrino oscillations.  On the other hand, the mixing 
of $\tilde \tau_L^c$ with $h^+$, i.e. $\xi$, should be of order $10^{-3}$ 
which is too large to be consistent with the usual terms of soft 
supersymmetry breaking.  For successful leptogenesis, 
the nonholomorphic term $H_2^\dagger H_1 \tilde \tau_L^c$ is required. 

\begin{figure}[htb]
\mbox{}
\vskip 3.6in\relax\noindent\hskip-0.4in\relax
\includegraphics{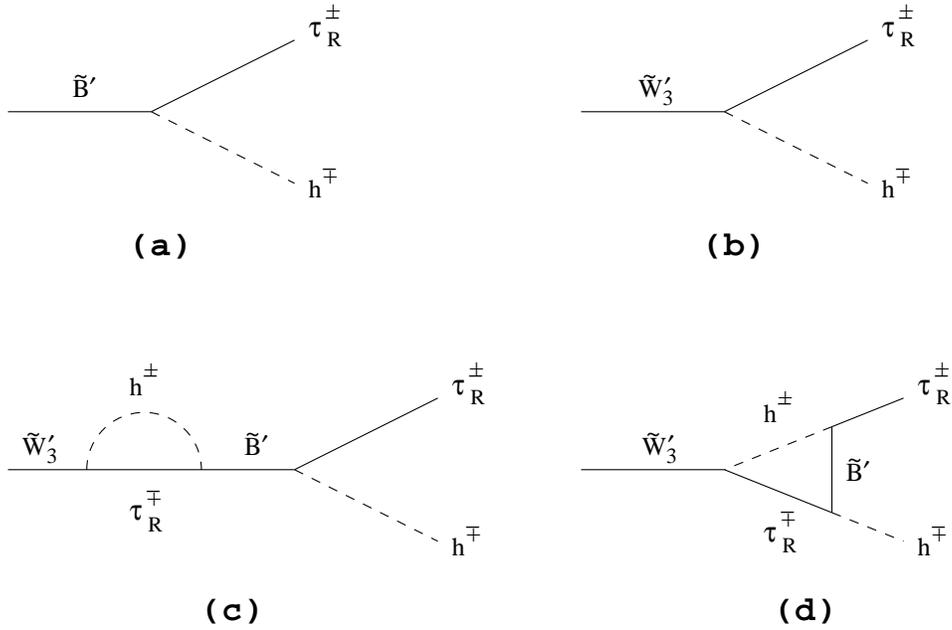} 
\vskip .25in
\caption{ Tree-level diagrams for (a) $\tilde B'$ 
decay and (b) $\tilde W'_3$ decay (through their $\tilde B$ content), 
and the one-loop (c) self-energy and (d) vertex diagrams for  
$\tilde W'_3$ decay which have absorptive parts of opposite lepton number.}
\end{figure}

\section{Conclusion and Outlook}

Models of neutrino mass and mixing invariably lead to other possible physical 
consequences which are important for our overall understanding of the 
Universe, as well as other possible experimentally verifiable predictions.

\section*{Acknowledgments}
I thank Rahul Basu and the other organizers of WHEPP-6 for their great 
hospitality and a stimulating meeting.  This work was supported in part by 
the U.~S.~Department of Energy under Grant No. DE-FG03-94ER40837.

\end{document}